\documentclass[aps, prl, twocolumn, superscriptaddress, longbibliography]{revtex4-2}

\usepackage{graphicx}
\usepackage{dcolumn}
\usepackage{bm}
\usepackage{todonotes}
\usepackage{color}
\usepackage{glossaries}
\usepackage{sidecap} 
\usepackage[colorlinks = true,
            linkcolor = red,
            urlcolor  = red,
            citecolor = blue,
            anchorcolor = blue]{hyperref}

\begin{document}

\title{Cooper-Pair Localization in the Magnetic Dynamics of a Cuprate Ladder}
\author{A.~Scheie}
\thanks{These authors contributed equally}
\affiliation{MPA-Q, Los Alamos National Laboratory, Los Alamos, New Mexico 87545, USA}

\author{P.~Laurell}
\thanks{These authors contributed equally}
\affiliation{Department of Physics \& Astronomy, University of Tennessee, Knoxville, Tennessee, USA}

\author{J.~Thomas}
\thanks{These authors contributed equally}
\affiliation{Department of Physics \& Astronomy, University of Tennessee, Knoxville, Tennessee, USA}
\affiliation{Institute of Advanced Materials and Manufacturing, University of Tennessee, Knoxville, Tennessee, USA}

\author{V.~Sharma}
\thanks{These authors contributed equally}
\affiliation{Department of Physics \& Astronomy, University of Tennessee, Knoxville, Tennessee, USA}

\author{A.~I.~Kolesnikov}
\affiliation{Neutron Scattering Division, Oak Ridge National Laboratory, Oak Ridge, Tennessee 37831, USA}

\author{G.~E.~Granroth}
\affiliation{Neutron Scattering Division, Oak Ridge National Laboratory, Oak Ridge, Tennessee 37831, USA}

\author{Q.~Zhang}
\affiliation{Neutron Scattering Division, Oak Ridge National Laboratory, Oak Ridge, Tennessee 37831, USA}

\author{B.~Lake}
\affiliation{Helmholtz Zentrum Berlin f{\"u}r Materialien und Energie GmbH, Hahn-Meitner Platz 1,D-14109 Berlin, Germany}
\affiliation{Institute for Solid State Physics, Technical University Berlin, Hardenbergstr. 36,10623 Berlin, Germany}

\author{M.~Mihalik~Jr.}
\affiliation{Institute of Experimental Physics, Slovak Academy of Sciences, Watsonova 47, Košice, Slovak Republic} 

\author{R.~I.~Bewley}
\affiliation{ISIS Neutron and Muon Source, Rutherford Appleton Laboratory, Didcot, OX11 0QX, United Kingdom}

\author{R.~S.~Eccleston}
\affiliation{ISIS Neutron and Muon Source, Rutherford Appleton Laboratory, Didcot, OX11 0QX, United Kingdom} 

\author{J.~Akimitsu}
\affiliation{Department of Physics, Aoyama-Gakuin University, Chitosedai, Setagaya-ku, Tokyo 157, Japan} 

\author{E.~Dagotto}
\affiliation{Department of Physics \& Astronomy, University of Tennessee, Knoxville, Tennessee, USA}
\affiliation{Materials Science and Technology Division, Oak Ridge National Laboratory, Oak Ridge, Tennessee 37831, USA} 

\author{C.~D.~Batista}
\affiliation{Department of Physics \& Astronomy, University of Tennessee, Knoxville, Tennessee, USA}
\affiliation{Shull Wollan Center, Oak Ridge National Laboratory, Tennessee 37831, USA}

\author{G.~Alvarez}
\affiliation{Computational Sciences and Engineering Division, Oak Ridge National Laboratory, Oak Ridge, Tennessee, USA}

\author{S.~Johnston}
\affiliation{Department of Physics \& Astronomy, University of Tennessee, Knoxville, Tennessee, USA}
\affiliation{Institute of Advanced Materials and Manufacturing, University of Tennessee, Knoxville, Tennessee, USA}

\author{D.~A.~Tennant}
\email[E-mail: ]{dtennant@utk.edu}
\affiliation{Department of Physics \& Astronomy, University of Tennessee, Knoxville, Tennessee, USA}
\affiliation{Shull Wollan Center, Oak Ridge National Laboratory, Tennessee 37831, USA}
\affiliation{Department of Materials Science and Engineering, University of Tennessee, Knoxville, Tennessee, USA}

\date{\today}

\begin{abstract}
We investigate the spin dynamics of the cuprate ladder  Sr$_{2.5}$Ca$_{11.5}$Cu$_{24}$O$_{41}$ to elucidate the behavior of its intrinsically doped holes. Combining high-resolution neutron spectroscopy and density matrix renormalization group calculations enables a comprehensive analysis of the collective magnetic dynamics. 
We find a general absence of magnetic signatures from unpaired charges, indicating holes within the system form strongly bound localized Cooper pairs. A one-band Hubbard model fails to match the spectral features but a straightforward extension to a large attractive nearest-neighbor interaction
quantitatively explains our results. Our finding shows the significance of additional interactions beyond the long-predicted quantum spin pairing in the ($d$-wave) charge pairing process.  Considering the parallels between ladders and two-dimensional cuprates, these results are potentially relevant for square lattices as well.
\end{abstract}

\maketitle

Cuprate ladders were introduced in the 1990's as stripped-down models of high-temperature (high-$T_\mathrm{c}$) superconductivity \cite{Dagotto1996}. These materials share many similarities with the two-dimensional cuprates, including superconductivity under pressure, yet are more amenable to theoretical modeling. Ladder behavior has also been invoked as a key component of recent theories of hole-doped planar cuprates based on intertwined stripes and pair-density-waves \cite{kivelson2020,Tranquada2021}. 
However, the details of the underlying pairing mechanism in cuprate ladders remains an open question. While the general belief is that strong electronic correlations play a crucial role, research on doped cuprate ladders has generated perplexing and contradictory outcomes when comparing theory and experiments, to the extent that even the doping levels of these materials have been a subject of controversy~\cite{Vuletic2006spinladder}. 

\begin{figure*}[t]
	\centering
    \includegraphics[width=\textwidth]{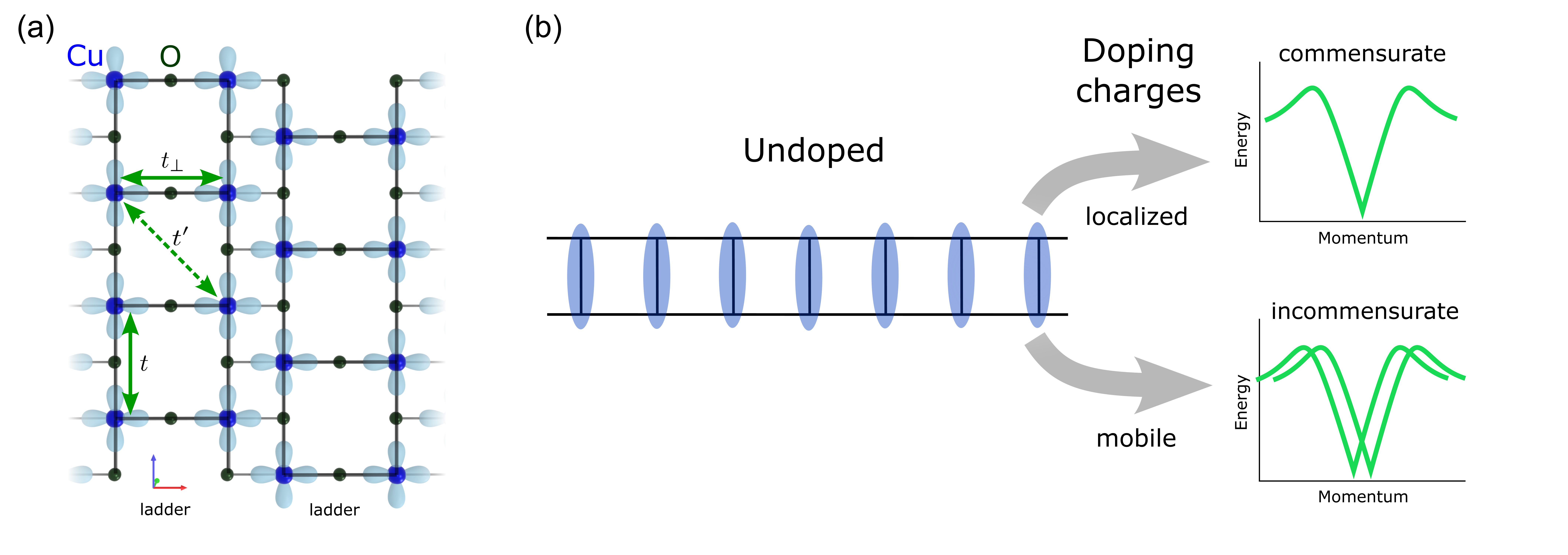}
	\caption{Ladder structures and magnetic dynamics of cuprate ladder Sr$_{2.5}$Ca$_{11.5}$Cu$_{24}$O$_{41}$. (a) A sketch of the ladder structure with the copper $3d_{x^2-y^2}$ and oxygen $2p_\sigma$ orbitals. The strong hybridization between the Cu and O orbitals leads to a strong superexchange coupling along both the ladder legs and rungs but weak coupling between the ladders. In a down-folded single-band Hubbard ladder description, these orbitals lead to hopping along the leg ($t$), rung ($t_\perp$), and diagonal direction ($t^\prime$), as indicated. (b) The expected one-triplon magnetic excitations for an undoped ladder with spin singlets on all rungs. When mobile charges are doped into the system, simulations show that they result in incommensurate split dispersions. As is shown below, the experimental Sr$_{2.5}$Ca$_{11.5}$Cu$_{24}$O$_{41}$ spectrum more closely resembles the undoped case, despite the fact that the ladder subsystem is doped.}
	\label{flo:Lattice}
\end{figure*}

An essential challenge in understanding these materials is to identify minimal Hubbard Hamiltonians that capture the high energy excitations associated with the underlying bare interactions, as well as the low-energy collective modes. The measured structure of the magnetic excitations thus imposes a stringent test for candidate models. Motivated by this, here we reexamine the case of Sr$_{2.5}$Ca$_{11.5}$Cu$_{24}$O$_{41}$ (SCCO). This material has long been the focal point of research as its magnetism and strong correlations are believed to be responsible for the formation of a possible charge-density-wave (CDW)~\cite{Abbamonte2004} and pressure-induced superconductivity~\cite{Nagata_1998}. Here we undertake high resolution inelastic neutron scattering (INS) measurements with matching density matrix renormalization group (DMRG) \cite{PhysRevLett.69.2863} calculations of the dynamical spin structure factor of an extended single-band Hubbard model. Through this, we pinpoint the principal interactions of a low-energy effective model for this prototypical ladder system and shed new light on the behavior of its doped holes.  

The SCCO family comprises Cu$_2$O$_3$ ladder layers separated by layers of Sr/Ca ions and CuO$_2$ chains, which serve as charge reservoirs. The Cu$_2$O$_3$ planes form a trellis lattice of spin-1/2 Cu sites, as shown in Fig.~\ref{flo:Lattice}a~\cite{Dagotto_1999}. 
Here, a subset of Cu ions are connected by 180$^{\circ}$ bonds to form two-leg ladders, with strong super-exchange interactions mediated by O ions. Neighboring ladders in the plane are displaced by $c/2$ ($c$ is the Cu-Cu distance along the leg) and connected by $90^{\circ}$ bonds that are also geometrically frustrated. This structure leads to a weak effective inter-ladder coupling such that the material's low-energy sector is dominated by contributions from largely independent ladder subsystems~\cite{Notbohm_2007, Schlappa2009collective}. 

In the absence of doping (e.g., in La$_4$Sr$_{10}$Cu$_{24}$O$_{41}$), these systems are Mott-Hubbard insulators in a quantum cooperative-paramagnetic state~\cite{Notbohm_2007} consisting of a resonant superposition of singlet spin-pairs with short-range entanglement and gapped triplon modes (see Fig. \ref{flo:Lattice}b). This physics is most easily understood by examining the strong rung coupling limit, where the dominant Heisenberg exchange coupling $J_\perp$ leads to a ground state with dominant spin singlet character on each rung~\cite{Dagotto1996,Troyer1996}. A finite energy $\Delta_T$ is required to break such a singlet and form a triplet (T) excited state carrying spin $S=1$ and charge 0 (i.e. a triplon quasi-particle). For cuprate ladders like SCCO, the leg and rung couplings are comparable, $J\approx J_\perp$, and the ground state includes pairs of singlets that resonate between different configurations to decrease energy. 
All excitations can be classified by the parity with respect to reflections about the centerline of the ladder: states with an odd (even) number of triplons have odd (even) parity. Consequently, the INS intensity of one-triplon excitations, associated with a $q_{\perp}=\pi/c$, wavevector, is modulated as $0.5[1-\cos(q_\perp c)]$, while the two-triplon scattering intensity, associated with the $q_{\perp}=0$ wavevector, is modulated as $0.5[1+\cos(q_\perp c)]$ ~\cite{Notbohm_2007}. INS measurements on the undoped ladder material La$_4$Sr$_{10}$Cu$_{24}$O$_{41}$ confirm the expected singlet ground-state, gapped triplon excitations, and parity behavior~\cite{Notbohm_2007}. The triplon dispersion is characterised by a highly dispersive mode with an energy gap $\Delta_T \approx 30$~meV, which is 
lower than the spin gap expected for a nearest-neighbor spin Hamiltonian. This reduction in the spin gap is due to the cyclic exchange $J_\mathrm{cyc}$ and further neighbor exchange interactions arising from significant charge fluctuations. (The ratio between the on-site Hubbard repulsion $U$ and the dominant hopping amplitude is $U/t \simeq 8$.) Indeed, a quantum phase transition is predicted for a strong enough cyclic exchange ($J_\mathrm{cyc}\approx J_\perp/3$), which has 
been observed in the related compound CaCu$_2$O$_3$ \cite{Lake2010}. 

The SCCO family is intrinsically charge doped with a formal valence on the copper sites of +2.25. While the majority of the 0.25 holes per Cu site reside in the chains, some holes are present in the ladder subsystem even without Ca doping. Although substitution of Ca for Sr does not alter the formal valence, the associated chemical pressure transfers additional holes to the ladders from the chains. However, the precise amount of charge transfer remains debated~\cite{Vuletic2006spinladder}. 
Nevertheless, different experimental approaches consistently place the ladder doping in the range $\delta=0.06-0.21$ for the heavily Ca doped systems. Moreover, Raman scattering \cite{Blumberg584,Gozar_2003}, electrical conductivity \cite{Gorshunov_2002, Blumberg584}, and x-ray diffraction measurements \cite{Abbamonte2004, Rusydi_2006} have all conclusively shown ladder charge order across Ca concentrations with superconductivity ($x=11-13$), consistent with the presence of doped holes in the ladder subsystem.

\begin{figure*}[ht]
    \centering
    \includegraphics[width=\textwidth]{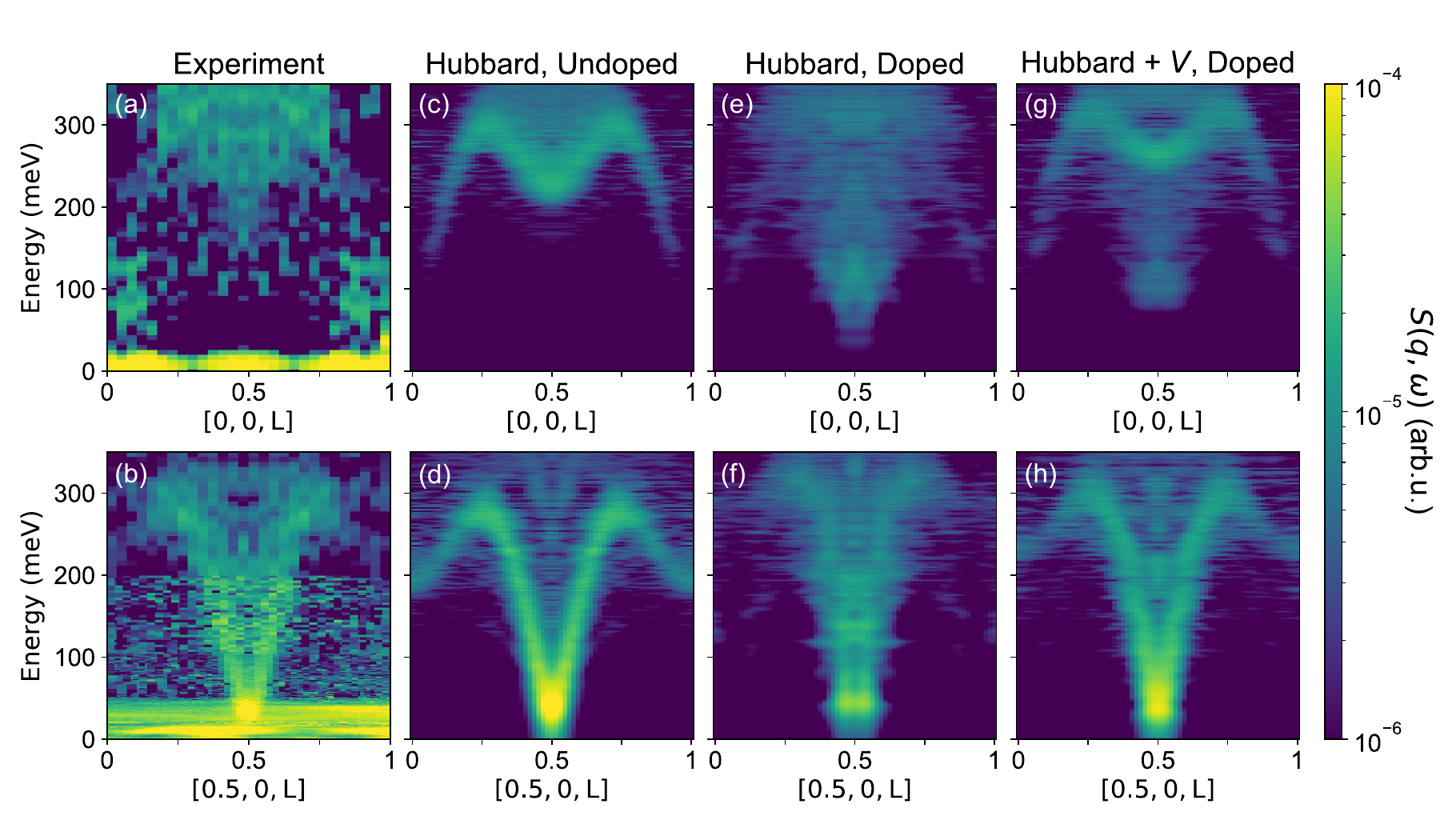}
    \caption{A comparison of the measured spin excitation spectrum and predictions from an extended Hubbard model. (a), (b) The measured INS spectra along the ${\bf q}=(q_\perp,0,L)$ direction for $q_\perp = 0$ and $0.5$ (r.l.u.), respectively, where the lattice spacing has been taken to be the Cu-Cu distance along the ladder leg. 
    (c) and (d) The dynamical spin structure factor $S({\bf q},\omega)$ obtained from DMRG calculations for an  undoped Hubbard Hamiltonian without any nearest-neighbor interactions ($V = 0$). (e), (f) DMRG results for $S({\bf q},\omega)$ obtained for a $\delta = 6.25\%$ hole-doped Hubbard Hamiltonian, again with $V = 0$. 
    (g), (h) $S({\bf q},\omega)$ for a $\delta = 6.25\%$ hole-doped Hubbard Hamiltonian with an attractive nearest-neighbor interaction $V = -t$. Note, the theory spectra shown in panels c-e and d-h have been convoluted with a Gaussian lineshape to reflect the experimental resolution (see methods). 
    }
    \label{flo:HighEnergy}
\end{figure*}

Hubbard ladder calculations show that doped holes form pairs~\cite{Troyer1996, Dagotto_1999, Dolfi2015pair, zhou2023robust}, which will either exhibit charge ordering or enter a superconducting state, depending on the interaction details. Doped holes will also affect the low-energy magnetic response, as each generates an anti-phase boundary in the antiferromagnetic correlations \cite{Broholm2000}. 
This leads to incommensurate magnetic fluctuations
at $q_\parallel= \pi (1 \pm \delta)/c$ ($q_\perp = \pi/c$), where $\delta$ is the effective hole doping of the ladder, as shown in Fig.~\ref{flo:Lattice}b. (Such incommensurate magnetic fluctuations are intrinsic to the doped AFM state, and are also observed in the magnetic excitations of doped $pd$-models for the cuprates~\cite{Li2021particle}.) However, as demonstrated in this study, we observe a remarkable absence of these incommensurate fluctuations in the INS data, despite the presence of doped holes in the ladder subsystem. We will show that this observation implies the presence of a strong Cooper pair localization that preserves the dominant commensurate fluctuations at $q_\parallel= \pi/c$ via  annihilation of opposite anti-phase boundaries. The presence of a stronger pairing than the one resulting from the bare repulsive Hubbard interaction suggests that other effects, such as ionic displacements, play a decisive role in the stabilization of high-$T_\mathrm{c}$ superconductivity.\\

We measured the neutron scattering spectrum of $\rm Sr_{2.5} Ca_{11.5} Cu_{24} O_{41}$ using the SEQUOIA spectrometer \cite{Granroth2006,Granroth2010}
at the ORNL SNS \cite{mason2006spallation} at $T=5$ and $120$~K, below and above the charge ordering temperature $T_\mathrm{CO}\sim80$~K \cite{Dagotto_1999}. Details about our sample  and preparation and INS measurements are provided in the Methods section.

Figs.~\ref{flo:HighEnergy}a and \ref{flo:HighEnergy}b show the dynamical spin structure factor $S(\bm{q},\omega)$ at 5~K for $q_{\perp}=0$ and $q_{\perp}=\pi/c$, respectively, after the data has been treated for background and magnetic form factor. To obtain a detailed image of the magnetic scattering, we overlay data from several incident energies to enhance the visibility of low-energy features. For an undoped ladder, these spectra probe two- and one-triplon scattering processes, respectively. The resolution into two parity components modulated by a cosine form factor in $q_{\perp}$ is also evident in the spectra, indicating that parity largely remains a good quantum number for $x=11.5$. 

To model SCCO's magnetic response, we performed DMRG simulations of an effective single-band extended Hubbard Hamiltonian (see also Methods)
\begin{equation}
    \mathcal{H}=-\sum_{i,j,\sigma}t_{ij}\left(c_{i,\sigma}^{\dagger}c^{\phantom\dagger}_{j,\sigma}+ {\rm H.c.}\right)+U\sum_{i}n_{i,\uparrow}n_{i,\downarrow}
    \label{eq:Hubbard_model}+ V \sum_{\langle i, j \rangle}n_{i}n_{j}, 
\end{equation}
where $c_{i\sigma}^{\dagger}$ ($c_{i\sigma}$) creates (annihilates) a spin-$\sigma$ electron on site $i$, $n_{i,\sigma} = c_{i,\sigma}^{\dagger}c^{\phantom\dagger}_{i,\sigma}$ is the spin-resolved number operator, $n_i = \sum_\sigma n_{i,\sigma}$, 
$t_{ij}$ are the hopping amplitudes between sites $i$ and $j$, and $U$ ($V$) is the on-site (nearest-neighbor) density-density interaction. Throughout, we restrict the hopping $t_{ij} \equiv t = 400$ meV for nearest-neighbor hopping along the legs, $t_{ij} \equiv t_\perp = 0.84t$ for hopping along the rungs, and $t_{ij} \equiv t^\prime = -0.3t$ for next-nearest-neighbor hopping along the diagonal. Further, we vary $U$ and $V$ as indicated throughout the main text such that the exchange interactions $J = 4t^2/(U-V)$ remain fixed. These parameter choices are identical to those also used in a recent resonant inelastic x-ray scattering study on the parent compound Sr$_{14}$Cu$_{24}$O$_{41}$~\cite{Padma2023}.

The predicted $S({\bf q},\omega)$ for the undoped Hubbard ladder with $U = 8t$ and $V = 0$ are shown in Figs.~\ref{flo:HighEnergy}c and \ref{flo:HighEnergy}d. 
If we instead assume a hole doping of $\delta=0.0625$ (on the lower end of the experimental estimates) we obtain the spectra in Figs.~\ref{flo:HighEnergy}e and \ref{flo:HighEnergy}f, with the expected low-energy incommensurate spin fluctuations featuring a considerably reduced gap. 
Surprisingly, the measured spectra show a striking degree of similarity with the undoped model, despite the expected hole doping of the Ca doped ladder. 

\begin{figure*}[t]
	\centering\includegraphics[width=\textwidth]{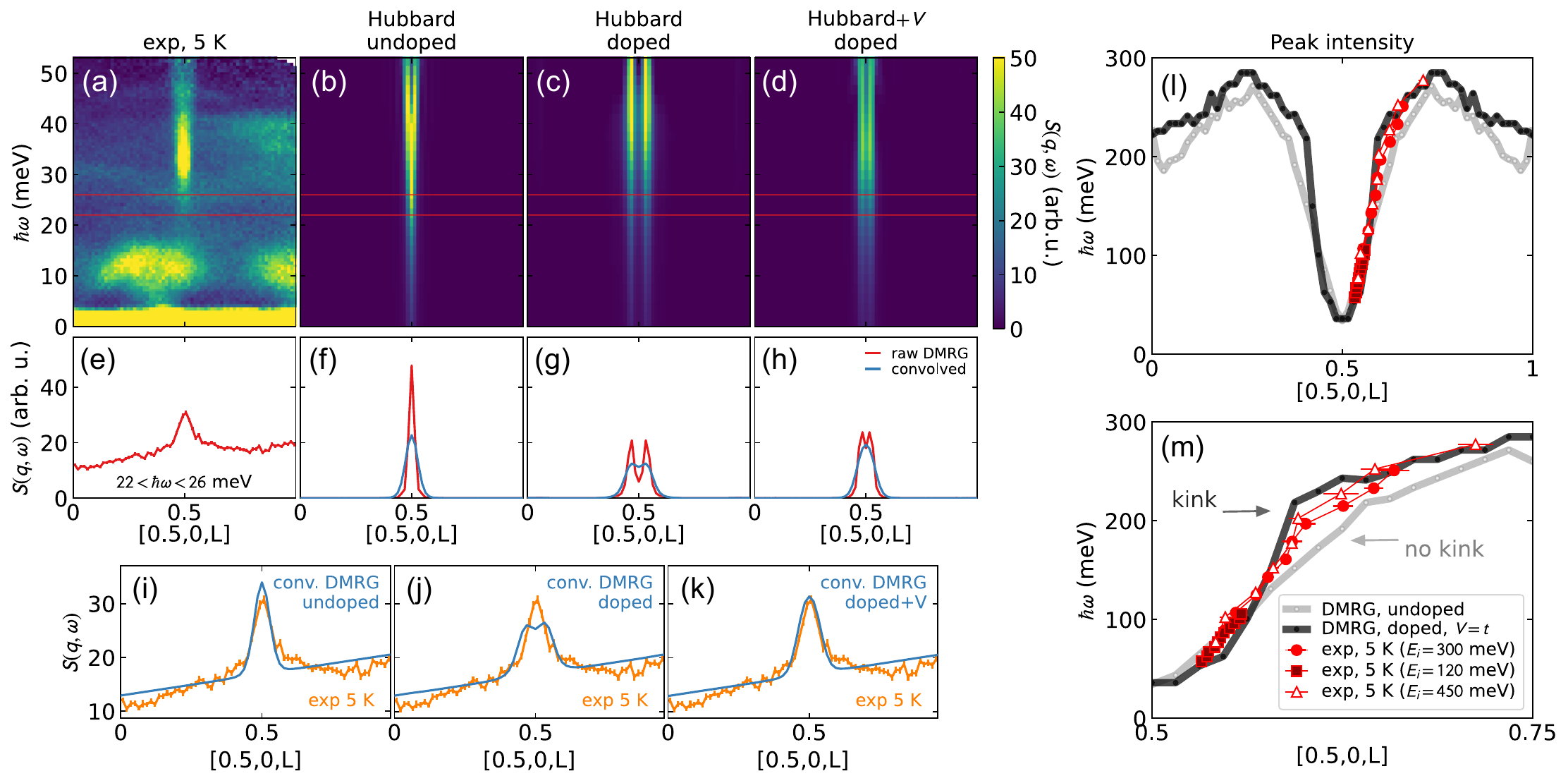}
	\caption{Lack of incommensuration in the low energy scattering. (a) the experimental $[\pi,0,L]$ scattering at 5~K. (b)-(d) DMRG calculated scattering for the undoped, doped, and extended Hubbard models, respectively. (e)-(h) cuts of the above panels between 22~meV and 26~meV (shown by the horizontal red lines).  According to DMRG simulations, as the doping increases the peak at $L=\pi$ splits to become incommensurate, but the addition of attractive interactions $V$ causes the peaks to come together again. Convolved with the experimental resolution (blue lines in panels (i)-(k), the slightly incommensurate peaks in the Hubbard$+V$ model become a single broad peak. As shown in panels i-k, the experimental scattering is resolution limited commensurate at $L=\pi$, indicating no observable incommensurability. (l), (m) the fitted dispersion of magnetic excitations for $q_{\perp}=\pi$ (red circles and line). DMRG results for undoped (gray lines) and the doped Hubbard$+V$ model (black lines) are also shown. The kink in the dispersion at $q_{\parallel}=\pi (1 \pm 0.125)/c$ is reproduced by the doped Hubbard$+V$ model but not the undoped Hubbard model model. }
	\label{flo:LowEnergy}
\end{figure*}

Our results, at first glance, suggest that the SCCO ladder system has a much lower effective doping than inferred from prior experiments. There are, however, notable differences between the INS data for SCCO and undoped Hubbard model predictions. First, the wavevector broadening of the dispersive magnetic excitations $q_{\perp}=\pi$ remains wider than resolution and becomes even more pronounced at higher energies. Second, there is a kink-like structure in the triplon dispersion, as shown in more detail in Figs.~\ref{flo:LowEnergy}l and \ref{flo:LowEnergy}m. This dispersion kink is absent in the measured response of the related undoped ladder material Lac$_4$Sr$_{10}$Cu$_{24}$O$_{41}$~\cite{Notbohm_2007} and very weak in the calculated spectra for the undoped Hubbard model. Third, the measured energy gap is not resolution limited and exhibits broadening, with subgap states extending down to $\sim 5$ meV. These states are sharp in $q$ and are centered at the commensurate wave vector $q_{\parallel}=\pi/c$, with no resolved incommensurate peaks.

The discrepancies between theory and experiment are deeply perplexing. It is well documented that doping the Hubbard ladder ($V=0$) leads to a Luther Emery liquid with a gapless charge mode and gapped spin excitations ($\Delta_s\simeq14$~meV for $\delta =0.0625$), whose  holes form pairs with a binding energy $|\Delta_B|\simeq 33$~meV for $\delta =0.0625$ (see SM). 
For small doping and $V=0$ the superconducting state is strongly favored; however, 
charge order and superconductivity compete once the doping gets to around 10\% \cite{Dolfi_2015}. 
Nonetheless, the crucial aspect of the ground state, which makes spin fluctuations highly sensitive to doping, is the minimization of hole kinetic energy when the spins located on both sides along the leg direction are anti-aligned. In other words, each doped hole in the Hubbard ladder carries an anti-phase boundary of the staggered magnetization favored by the effective antiferromagnetic Heisenberg interactions~\cite{Batista00}. This induces a modulation of antiferromagnetic correlations with a wavelength equal to the average inter-hole distance that results in the dominant dynamical susceptibility occurring at incommensurate wave vectors $q_{\parallel} = \pi (1 \pm \delta)/c$. This is in clear contradiction to experiment.

The required annihilation of opposite anti-phase boundaries can be achieved by a stronger hole-hole attraction, which reduces the coherence length $\xi$ of each Cooper pair and reinstates dominant commensurate antiferromagnetic (AFM) fluctuations at $q_{\parallel} = \pi/c$. Such an increased attraction can be achieved by introducing a nearest-neighbor {\it attractive} interaction $V < 0$ of order the hopping $t$. Notably, such a strong attractive interaction has recently been inferred in doped 1D spin-chain cuprates~\cite{Chen2021anomalously}, although without the Cooper pairs observed here, and has been attributed to electron-phonon coupling~\cite{Chen2021anomalously, Wang2021phonon}. 
We find that reducing $\xi$ to the scale of a few lattice spacings necessitates a combination of an attractive potential with $|V|\approx t$, along with localization induced by pinning centers generated by impurities.
This behavior is consistent with the fact that SCCO is insulating at low temperatures, implying that indeed the holes are localized. We can simulate these effects by introducing attractive pinning centers, mimicking impurities or disorder in the actual material. Interestingly, we find that the presence of such pinning centers reinforces the role of $V$ by further reducing the coherence length $\xi$. The small density $\delta'$ of holes that remain in the bulk (with a localization length scale longer than the inter-impurity distance) are still expected to produce incommensurate peaks at wave vectors $q_{\parallel}= \pi (1 \pm \delta')/c$. For small enough $\delta'$, the incommensuration would become smaller than the experimental resolution, explaining the similarity with the spectrum of the undoped compound.

This proposed scenario can be directly corroborated by the DMRG simulations presented in Fig.~\ref{flo:HighEnergy}g and \ref{flo:HighEnergy}h, which introduce an attractive Hubbard interaction $V=-t$. In this case, the function of the pinning centers is emulated by employing open boundary conditions, which generate an attractive boundary potential for Cooper pairs when $V < 0$. As anticipated, the combination of the pinning potential at the boundaries and a sizable attractive interaction localizes a large number of Cooper pairs near the ends of the cluster, significantly suppressing the hole density across the remainder of the ladder. 
The splitting $\Delta q_{\parallel}= 2 \pi \delta/c$ of the former incommensurate peaks now reduces to $\Delta q_{\parallel}=2 \pi \delta'/c $  with a momentum space broadening caused by charge fluctuations. The comparison between Figs.~\ref{flo:HighEnergy}a and \ref{flo:HighEnergy}g also shows that the model reproduces the observed two-triplon continuum. 

While the peak in the spin intensity of measured ($\Delta_T\approx32$ meV) and calculated spectra (38 meV) are at comparable energies, very weak ``subgap'' scattering extends downwards from the  broadened gap at $\Delta_T$. 
Figs.~\ref{flo:LowEnergy}a-k examine the low-energy subgap states of both the measured and calculated magnetic responses. It is essential here to differentiate between ladder and chain signals. The scattering from the latter manifests as a discernible band around 10~meV due to their significantly lower superexchange interaction. Given the distinct lattice spacing of the chains of approximately $c/\sqrt{2}$, their scattering can also be identified by its lack of symmetry around $q_{\parallel}=\pi$. The lack of signal below the chain scattering points to even this small subgap signal having a quantum spin gap ($\Delta_\mathrm{subgap}\gtrsim 5$~meV). The momentum resolution is finite in these data, which makes it challenging to distinguish a single feature from two very close features. However, as Fig.~\ref{flo:LowEnergy}i-k shows, the resolution-convolved $S({\bf q},\omega)$ obtained from the doped Hubbard model without $V$ has a visible double-peak structure, which allows us to rule out this scenario. Meanwhile the best match is obtained by the Hubbard$+V$ model broadened by the experimental resolution. (We estimate momentum resolution using the lowest energy $120$~K data, assuming that the $L=\pi$ feature is resolution-limited. It is notable that at $120$~K the gap feature at $32$~meV strongly weakens and the scattering extends to zero energy as would be expected for thermal delocalization, see SM). The Hubbard+$V$ model also produces a better overall description of the higher-energy features at low temperature in the triplon dispersion, including both its bandwidth and its kink around $q_{\parallel} = \pi (1 \pm 0.125)/c$, as shown in Figs.~\ref{flo:LowEnergy}l and \ref{flo:LowEnergy}m. 
 
\begin{figure*}[t]
    \centering
    \includegraphics[width=\textwidth]{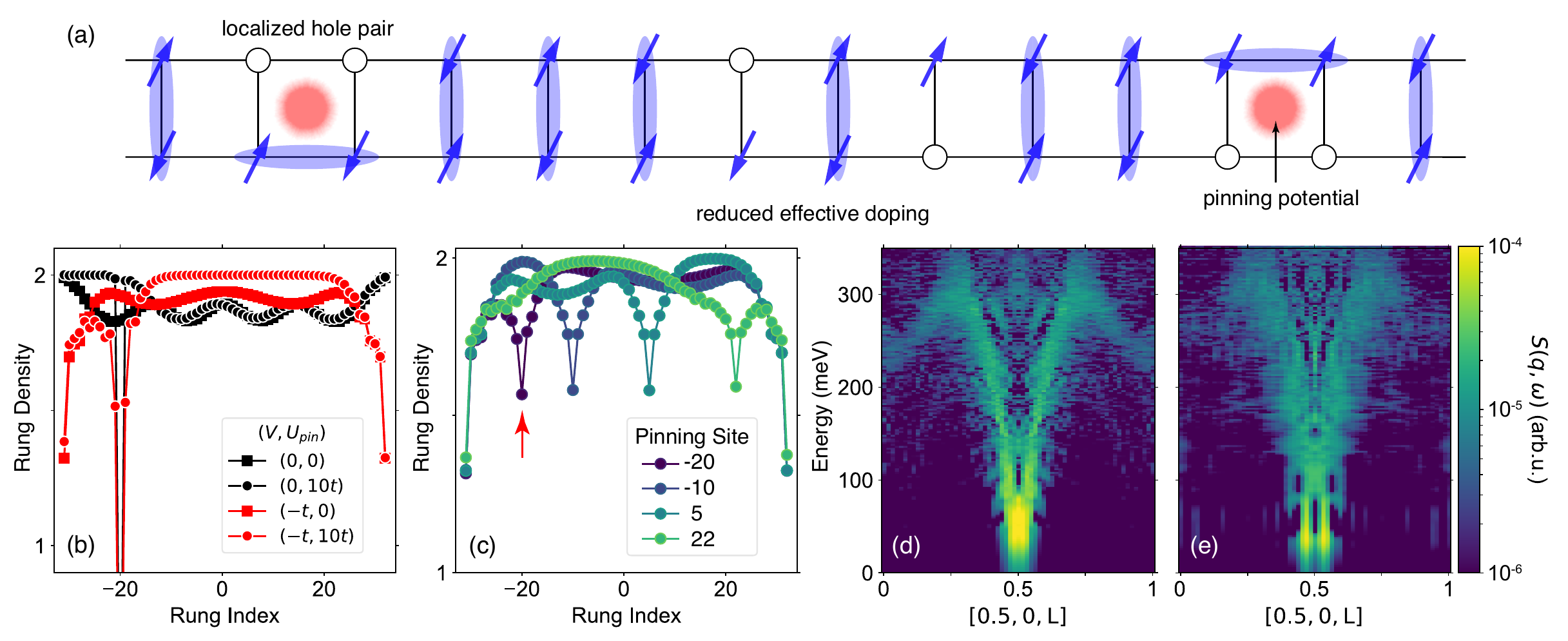}
    \caption{
        {\bf Figure 4: The combined role of enhanced pairing and impurities in reducing the effective doping.} {\bf a} The distribution of doped holes in the ladder. The attractive interaction $V$ enhances hole pairing, increases the effective mass of the carriers, and makes them prone to localization. Pair localization reduces the effective doping of the ladder and minimizes the disruption to the antiferromagnetic correlations. {\bf b} The electron rung density of a $64\times 2$ Hubbard ladder with a hole pinning potential $U_\mathrm{pin}$ placed on the 20\textsuperscript{th} rung. The holes are distributed uniformly throughout the cluster when $V =0$ and $U_\mathrm{pin} = 0$. When $V =-t$ but no potential is applied ($U_\mathrm{pin} = 0$), the open ends of the cluster act as pinning centers that reduce the carrier concentration in the center. When $V =-t$ and a strong $U_\mathrm{pin} = 10t$ is introduced, holes instead collect around both the impurity  potential and open boundaries. Finally, a strong $U_\mathrm{pin} = 10t$ \textit{without} $V$ localizes two holes on the rung. {\bf c} The rung electron densities for an extended Hubbard model with $V = -t$ and a weak $U_\mathrm{pin} = t/2$ placed on different rungs. Due to the large value of $V$, even a weak pinning potential is sufficient to trap many holes near the impurity sites. {\bf d} $S({\bf q},\omega)$ along $q_\perp = \pi$ for one of the cases shown in panel {\bf c}, as indicated by the red arrow. {\bf e} $S({\bf q},\omega)$ along $q_\perp = \pi$ for the $V=0$, $U_\mathrm{pin}=10t$ case shown in panel {\bf b}.
    }\label{fig:impurities}
\end{figure*}

Our results indicate that the magnetic response of the doped SCCO family can be understood in  a scenario where the doped holes experience enhanced Cooper-pair localization. To further validate this hypothesis, Fig.~\ref{fig:impurities} presents results obtained from simulations where the pinning centers are distributed across the ladder instead of being induced by the open boundary conditions of the cluster. Fig.~\ref{fig:impurities}a  provides a schematic representation of the scenario emerging from the results presented in the subsequent panels. Fig.~\ref{fig:impurities}b shows the rung densities of the ladder when a strong pinning potential $U_\mathrm{pin} = 10 t$ is displaced from the boundary (positioned on the 20\textsuperscript{th} rung). If $V = 0$, this pinning potential only captures two holes, leaving the remaining holes free in the bulk of the cluster. The resulting magnetic excitations, shown in Fig.~\ref{fig:impurities}e resembles that of a doped ladder and are inconsistent with our observations. Disorder averaging over a distribution of impurities does not improve the situation~\cite{tseng2022crossover}. Instead, if an attractive $V$ term is added, then tightly bound Cooper pairs are collected around the pinning center. This result shows that Cooper pair localization, amplified by an enhanced pair binding, is required to explain the observed experimental spectra. This amplification effect is attributed to the expected increase of the effective mass of the Cooper pair resulting from the reduction of the coherence length $\xi$. 

The results shown in Fig.~\ref{fig:impurities}c further confirm that the attractive $V$ term amplifies the localization produced by an attractive pinning center. Here, we place much weaker attractive potentials $U_\mathrm{pin} = t/2$ at different rungs of the ladder and find that each traps a similar amount of charge within its vicinity. Fig.~\ref{fig:impurities}d shows the dynamical spin structure factor $S({\bf q},\omega)$ along $q_{\perp} = \pi$ for one of the cases depicted in Fig.\ref{fig:impurities}c, which confirms this weak potential is sufficient to suppress the incommensurate fluctuations (see also SM). Direct comparison of \ref{flo:HighEnergy}d and Fig.~\ref{fig:impurities}d demonstrates that the attractive $V=-t$ interaction is crucial for explaining quantitatively the experimental data for the doped ladders using an effective single band model. We also note that moving to a three band description will not solve the problem as incommensurate peaks also appear in the magnetic spectrum of those models when doped~\cite{Li2021particle}. We have explicitly checked the dominant pairing correlations have a $d$-wave form when $V < 0$~\cite{Dagotto1996,Sigrist_1994}, consistent with other recent studies~\cite{jiang2022enhancing, Padma2023, zhou2023robust, Peng2023enhanced}.

Applying pressure may cause further delocalization explaining why pressure induces the superconducting transitions in Sr$_{2.5}$Ca$_{11.5}$Cu$_{24}$O$_{41}$. Further work on pressure effects and electron-phonon couplings in the ladder materials can be expected to shed new light on the detailed charge-pairing mechanisms and the superconducting transition, including how the incommensurate wavevector changes going into the superconducting state. \\

\section*{Summary and conclusions}
Our experimental findings suggest that cuprate ladders exhibit a tendency towards superconducting correlations, but experience localization of paired holes. A central conclusion drawn from our study is that an additional attractive interaction $V$ between nearest neighbor sites is essential to explain the minimal disruption observed in the dynamical spin correlations induced by hole doping. While the attractive $V$ term may be effectively mimicking the role of degrees of freedom that are not explicitly included in the Hubbard model (e.g.  electron-phonon coupling~\cite{Chen2021anomalously, Wang2021phonon}), it is important to highlight the crucial role of this term in favoring localization of small Cooper pairs around weak pinning centers by simultaneously reducing the coherence length $\xi$ and increasing the pair effective mass. 

Cuprate ladders, as long proposed, provide a fascinating platform to understand the interplay of charge and spin degrees of freedom as a stripped down model of high-temperature superconductivity, albeit with localization effects from their intrinsic composite structure.

\section*{Methods}
\noindent{\bf Sample Growth} --- Single crystals of Sr$_{14-x}$Ca$_x$Cu$_{24}$O$_{41}$ ($x=11.5$) (see also Supplementary Fig. 1) were grown in an image furnace using the traveling solvent method following the procedure described in Ref.~\cite{ammerahl1998}. As a starting material, we used CaCO$_3$ and SrCO$_3$ (purity better than 3N5) from Alpha Aesar and CuO (purity 4N) from Sigma Aldrich. Parts of all grown crystals were checked by X-ray powder diffraction and verified to have a single phase. The single-crystallinity of the sample was checked by E5 four-circle neutron diffractometer (Helmholtz Zentrum Berlin, Germany).  \\

\noindent {\bf Neutron experiments} --- We measured the inelastic neutron spectrum of $\rm Sr_{2.5} Ca_{11.5} Cu_{24} O_{41}$ using a 5 gram single crystal, mounted in the $0KL$ plane with the ladder legs ($c$ direction) extending in the horizontal plane and the ladder rungs ($a$ direction) extending in the vertical plane (Fig.~\ref{flo:Lattice}). We measured the spectra with incident energies of 30~meV, 60~meV, 120~meV, 300~meV, and 450~meV in a fixed orientation with the ladder legs perpendicular to the incident beam. The instrument settings for each energy configuration are given in Tbl.~1 below. Measurements were carried out at 5~K and 120~K, although the 450~meV setting was only measured at 5~K. The complete data set is shown in Supplementary Figure 2. 
\begin{table}[h]
    \centering
    \caption{Chopper frequencies for the various SEQUOIA settings, run in High-flux mode.}
    \begin{tabular}{c|cc}
        $E_i$ (meV) & $T0$ (Hz) & Fermi 1  (Hz) \\ 
        \hline
        30 & 60 & 120  \\
        60 & 90 & 180  \\
        120 & 90 & 240 \\
        300 & 120 & 420 \\
        450 & 150 & 480
    \end{tabular}
    \label{tab:chopper_settings}
\end{table}

The $\rm Sr_{2.5} Ca_{11.5} Cu_{24} O_{41}$ magnetic scattering is independent of the $k$ reciprocal space direction due to the decoupled ladder planes, so we can gather all the relevant information about the magnetic spectrum in the $H$ and $L$ directions from a fixed angle scan. To verify this, we also measured the inelastic spectra at $E_i=60$~meV rotating $180^{\circ}$, and found that the magnetic scattering did not vary with $K$ (see Supplementary Figure 3).

To determine the background, we also measured the scattering in the same configurations (incident energies 30, 60, 120, 300, and 450~meV) with the sample removed. We then smoothed this background data with a Gaussian profile to make up for smaller counting statistics and subtracted it from the sample scattering, shown in Supplementary Figure 4. After subtracting the background, we corrected for the anisotropic $d_{x^2-y^2}$ Cu$^{2+}$ form factor~\cite{zaliznyak2005magnetic} 
\begin{align*}
    f({\bf Q})=&\langle j_0 \rangle +
    \frac{5}{7}(3 \cos^2 \beta - 1) \langle j_2 \rangle \\
    &+ 
    \frac{3}{56}\big[35 \cos^4 \beta -  30 \cos^2 \beta \\
    &\quad\quad\quad+  35 \sin^4 \beta \cos 4 \alpha + 3\big]\langle j_4 \rangle, 
\end{align*}
where $\beta$ is the angle between $\bf Q$ and the $d_{x^2-y^2}$ orbital's $z$ axis (which in this case is the $b$ direction), and $\alpha$ is the $xy$-plane angle measured from the $x$ axis. The $\langle j_n \rangle$ constants for Cu$^{2+}$ were taken from Ref. \cite{BrownFF}. 
In this case, the difference between the isotropic and anisotropic form factor is dramatic at high energies because the $d_{x^2-y^2}$ orbitals order in the Cu ladder planes; see Supplementary Figure 5. After subtracting the background and correcting for the form factor, we were able to stitch together the five different incident energies to produce the plots shown in the main text Fig. 2. The overall intensities of the spectra taken at different energies were re-scaled by comparing constant-energy integrated intensity of the dispersive magnetic modes to ensure that the overall intensity is consistent across the different energy settings. \\

\noindent{\bf Density Matrix Renormalization Group calculations} --- 
Zero-temperature DMRG \cite{PhysRevLett.69.2863, PhysRevB.48.10345} calculations were performed using the DMRG++ software \cite{Alvarez2009} and open boundary conditions. We kept up to $m_\mathrm{max}=2000$ DMRG states on $64\times 2$ ladders for the ground state calculations. The dynamical spin structure factor $S({\bf q},\omega)$ was calculated using the Krylov correction-vector method \cite{PhysRevB.60.335, PhysRevB.66.045114, PhysRevE.94.053308} using the center-site approximation. When calculating $S({\bf q},\omega)$, we kept up to $m_\mathrm{max}=1000$ DMRG states and used a Lorentzian energy broadening with half width at half maximum equal to $\eta=0.03t$, and a frequency step $\Delta \omega = 0.005t$. All $S({\bf q},\omega)$ DMRG results have been uniformly scaled to match the experimental intensities. 

The doping estimate for SCCO varies between $6\%-20\%$ [see Refs.~\cite{Vuletic2006spinladder} and supplementary Figure 4 of Ref. \cite{tseng2022crossover}]. Supplementary Figure 6 shows the dynamical spin structure factor for various doping values in this range with $V=0$. Even for the lowest doping estimate of $6.25\%$, the incommensurate peaks are visible around $(\pi,\pi)$ in the DMRG simulations but absent in the experiment. The differences between theory and experiment only become worse with increased doping. We pick $6.25\%$, on the lower end of the estimated doping range, to obtain a conservative estimate of the magnitude of $V$.\\

\noindent{\bf Acknowledgements}: We acknowledge stimulating and illuminating discussions with Goetz Uhrig, Alexei Tsvelik, and Fabian Essler. We also are deeply grateful to Isabelle Exius and Susanne Notbohm for discussions and sharing results of their previous experiments. The work by D.~A.~T., S.~J., J.~T., V.~S., and C.~D.~B. were supported by the U.S. Department of Energy, Office of Science, Office of Basic Energy Sciences, under Award Number DE-SC0022311. The work by P.~L. and E.~D. was supported by the U.S. Department of Energy, Office of Science, Basic Energy Sciences, Materials Sciences and Engineering Division. A.~S. and G.~A. were supported by the U.S. Department of Energy, Office of Science, National Quantum Information Science Research Centers, Quantum Science Center. This research used resources at the Spallation Neutron Source, a DOE Office of Science User Facility operated by the Oak Ridge National Laboratory. Software development has been partially supported by the Center for Nanophase Materials Sciences, which is a DOE Office of Science User Facility.

\appendix

\bibliography{references}

\end{document}


\title{Supplementary Material for ``Cooper-Pair Localization in the Magnetic Dynamics of a Cuprate Ladder''}
\author{A.~Scheie}
\thanks{These authors contributed equally}
\affiliation{MPA-Q, Los Alamos National Laboratory, Los Alamos, New Mexico 87545, USA}

\author{P.~Laurell}
\thanks{These authors contributed equally}
\affiliation{Department of Physics \& Astronomy, University of Tennessee, Knoxville, Tennessee, USA}

\author{J.~Thomas}
\thanks{These authors contributed equally}
\affiliation{Department of Physics \& Astronomy, University of Tennessee, Knoxville, Tennessee, USA}
\affiliation{Institute of Advanced Materials and Manufacturing, University of Tennessee, Knoxville, Tennessee, USA}

\author{V.~Sharma}
\thanks{These authors contributed equally}
\affiliation{Department of Physics \& Astronomy, University of Tennessee, Knoxville, Tennessee, USA}

\author{A.~I.~Kolesnikov}
\affiliation{Neutron Scattering Division, Oak Ridge National Laboratory, Oak Ridge, Tennessee 37831, USA}

\author{G.~E.~Granroth}
\affiliation{Neutron Scattering Division, Oak Ridge National Laboratory, Oak Ridge, Tennessee 37831, USA}

\author{Q.~Zhang}
\affiliation{Neutron Scattering Division, Oak Ridge National Laboratory, Oak Ridge, Tennessee 37831, USA}

\author{B.~Lake}
\affiliation{Helmholtz Zentrum Berlin f{\"u}r Materialien und Energie GmbH, Hahn-Meitner Platz 1,D-14109 Berlin, Germany}
\affiliation{Institute for Solid State Physics, Technical University Berlin, Hardenbergstr. 36,10623 Berlin, Germany}

\author{M.~Mihalik~Jr.}
\affiliation{Institute of Experimental Physics, Slovak Academy of Sciences, Watsonova 47, Košice, Slovak Republic} 

\author{R.~I.~Bewley}
\affiliation{ISIS Neutron and Muon Source, Rutherford Appleton Laboratory, Didcot, OX11 0QX, United Kingdom}

\author{R.~S.~Eccleston}
\affiliation{ISIS Neutron and Muon Source, Rutherford Appleton Laboratory, Didcot, OX11 0QX, United Kingdom} 

\author{J.~Akimitsu}
\affiliation{Department of Physics, Aoyama-Gakuin University, Chitosedai, Setagaya-ku, Tokyo 157, Japan} 

\author{E.~Dagotto}
\affiliation{Department of Physics \& Astronomy, University of Tennessee, Knoxville, Tennessee, USA}
\affiliation{Materials Science and Technology Division, Oak Ridge National Laboratory, Oak Ridge, Tennessee 37831, USA} 

\author{C.~D.~Batista}
\affiliation{Department of Physics \& Astronomy, University of Tennessee, Knoxville, Tennessee, USA}
\affiliation{Shull Wollan Center, Oak Ridge National Laboratory, Tennessee 37831, USA}

\author{G.~Alvarez}
\affiliation{Computational Sciences and Engineering Division, Oak Ridge National Laboratory, Oak Ridge, Tennessee, USA}

\author{S.~Johnston}
\affiliation{Department of Physics \& Astronomy, University of Tennessee, Knoxville, Tennessee, USA}
\affiliation{Institute of Advanced Materials and Manufacturing, University of Tennessee, Knoxville, Tennessee, USA}

\author{D.~A.~Tennant}
\email[E-mail: ]{dtennant@utk.edu}
\affiliation{Department of Physics \& Astronomy, University of Tennessee, Knoxville, Tennessee, USA}
\affiliation{Shull Wollan Center, Oak Ridge National Laboratory, Tennessee 37831, USA}
\affiliation{Department of Materials Science and Engineering, University of Tennessee, Knoxville, Tennessee, USA}

\date{\today}

\maketitle

\begin{figure}
    \centering
    \includegraphics[width=0.5\textwidth]{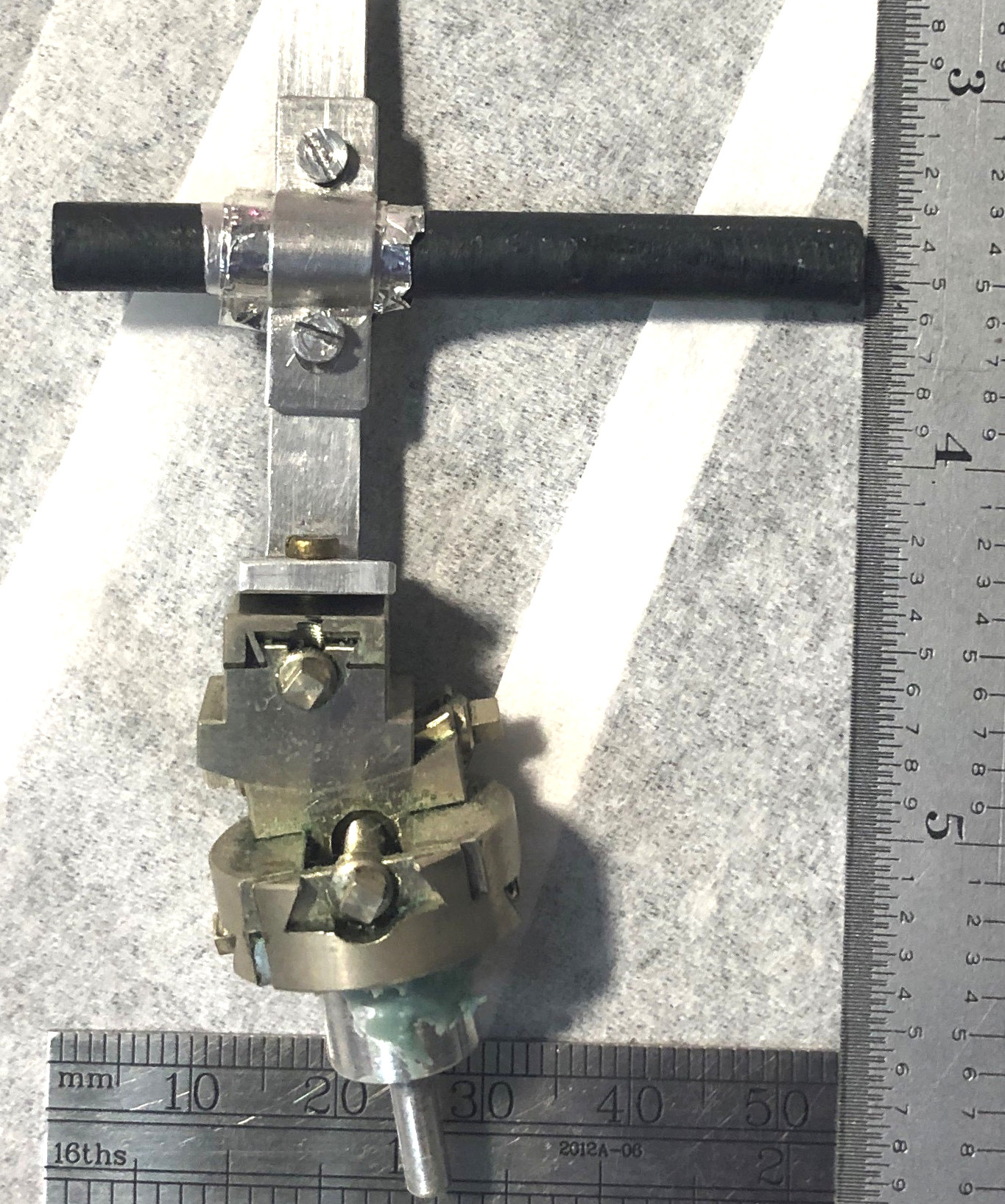}
    \caption{
    The single crystal of $\rm Sr_{2.5} Ca_{11.5} Cu_{24} O_{41}$ and sample holder used for neutron experiments. The ruler on the bottom marks mm, the ruler on the right marks inches.}
    \label{fig:crystal}
\end{figure}

\begin{figure}
    \centering
    \includegraphics[width=0.9\textwidth]{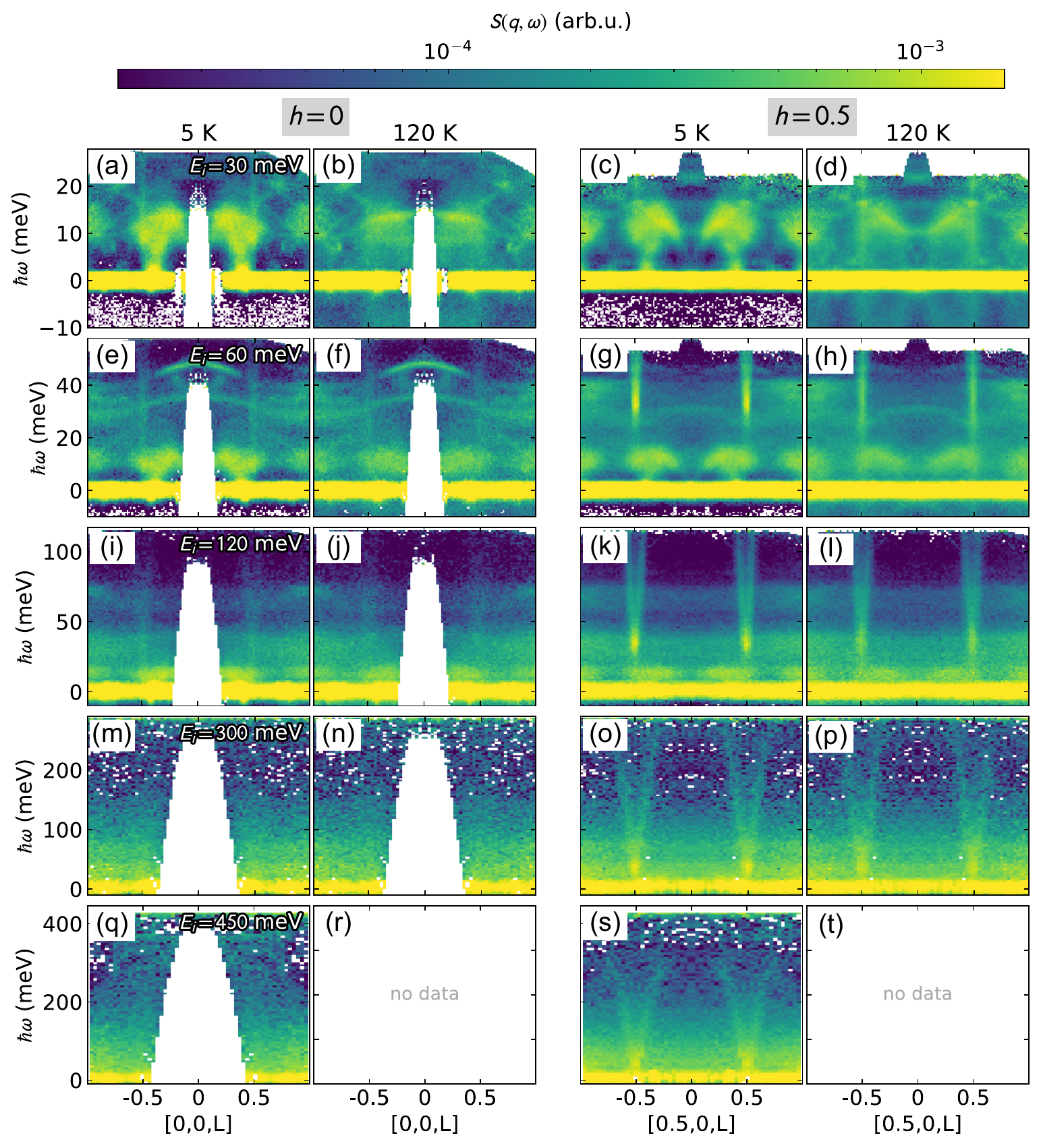}
    \caption{Sr$_{2.5}$Ca$_{11.5}$Cu$_{24}$O$_{41}$ scattering along chains collected with the incident energies in Table 1, at both $h=\pi$ and $h=0$.}
    \label{fig:FullData}
\end{figure}

\begin{figure}
    \centering
    \includegraphics[width=0.6\textwidth]{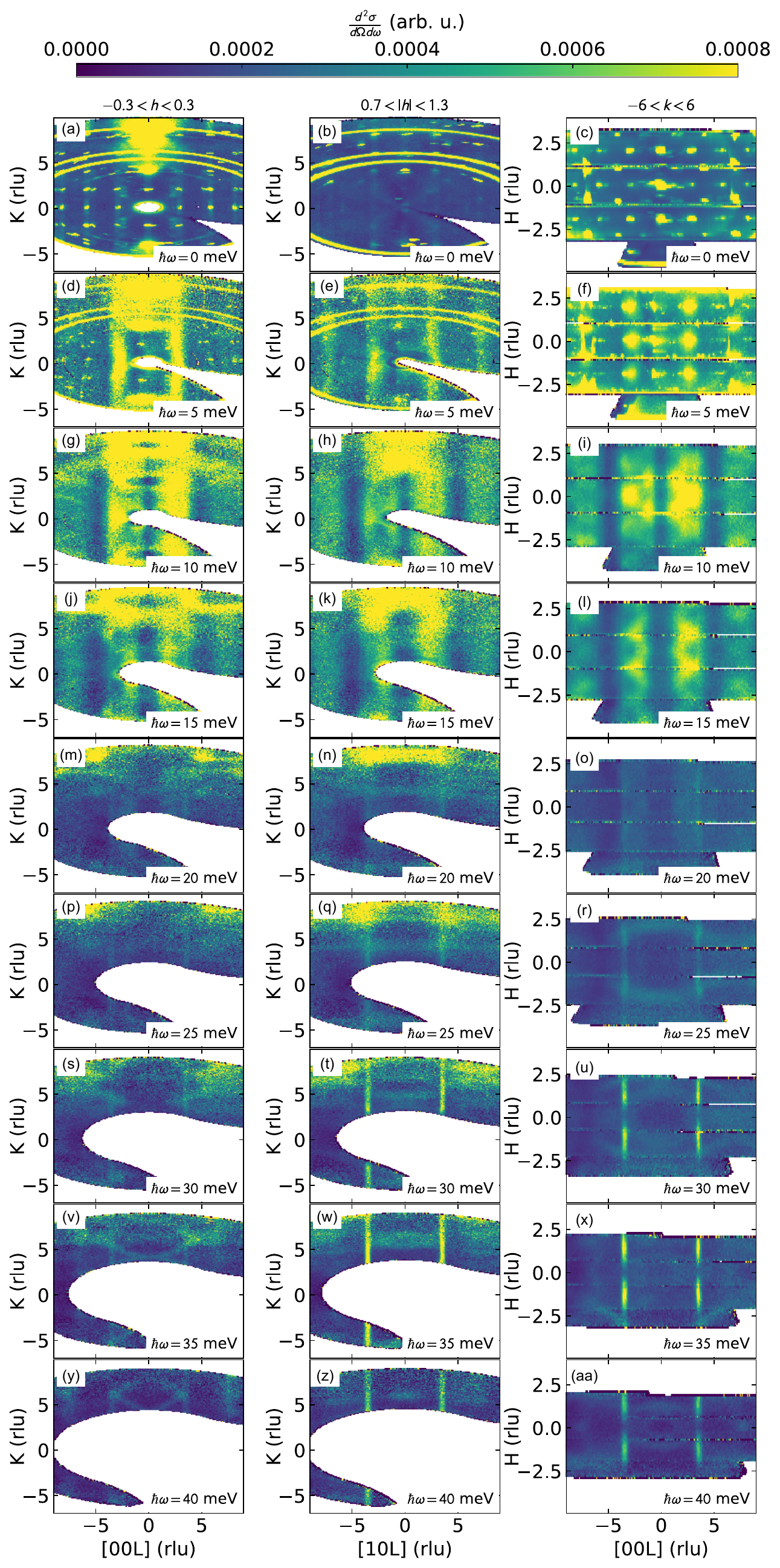}
    \caption{$T=5$~K Rotation scans of $\rm Sr_{2.5} Ca_{11.5} Cu_{24} O_{41}$ at $E_i = 60$~meV at incident energy transfers between $\hbar \omega = 0$ and $\hbar \omega = 40$~meV. As shown in the two left columns, the magnetic scattering is independent of $K$, which justifies using fixed-angle scans which (when corrected for the magnetic form factor) fully describe the magnetic spectrum. The right column shows the scattering along $H$ and $L$, revealing the short-range correlations between the ladder rungs. Note that, as opposed to the plots in the main text, the units here are in $\text{unit}~\text{cell}$ lattice units, such that $L_\text{unit~cell} = 7 L_\text{ladder}$ and $H_\text{unit~cell} = 2 H_\text{ladder}$.}
    \label{fig:Rotation5K}
\end{figure}

\begin{figure}
    \centering
    \includegraphics[width=0.8\textwidth]{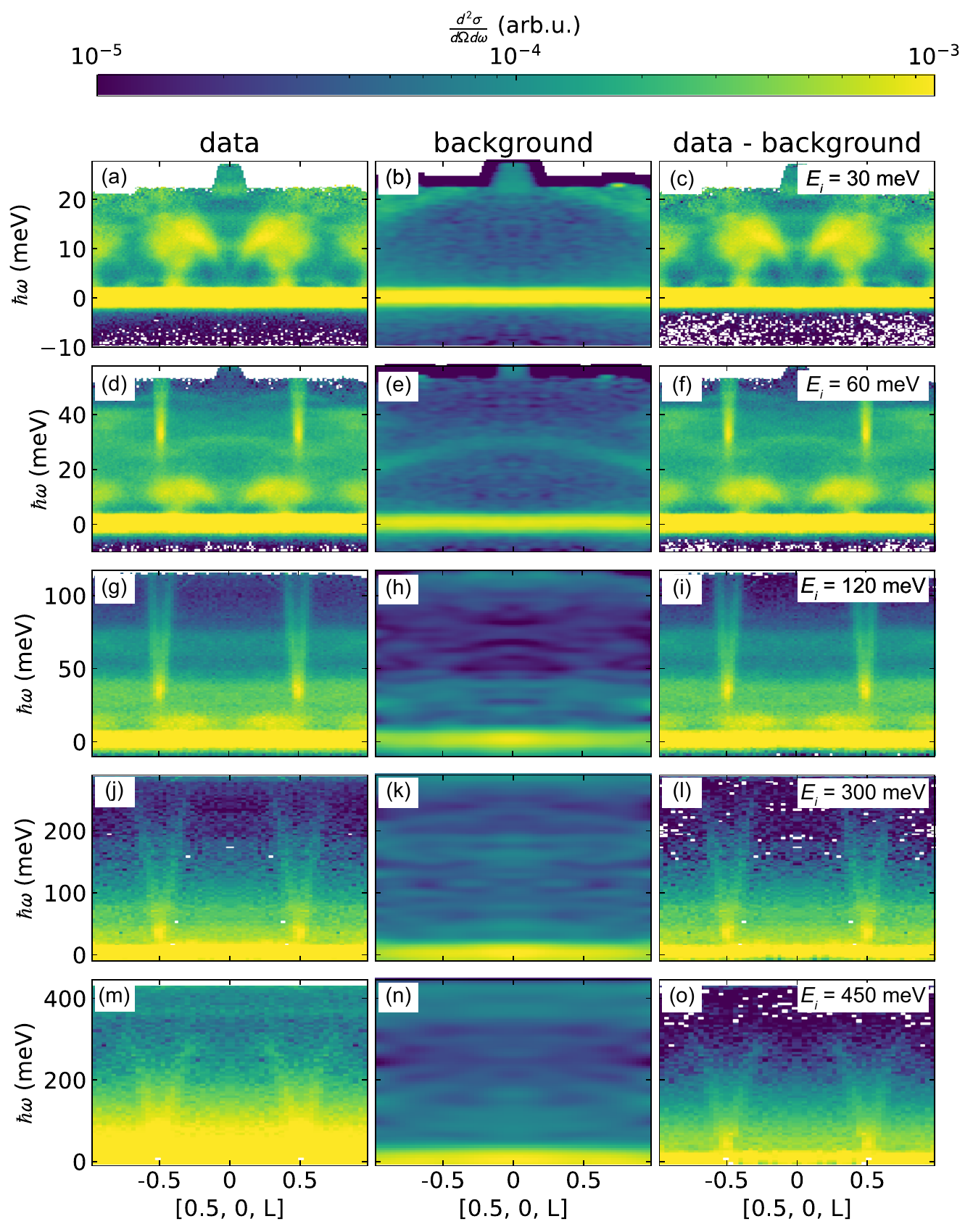}
    \caption{$\rm Sr_{2.5} Ca_{11.5} Cu_{24} O_{41}$ background subtraction routine, wherein a smoothed empty measurement (sample removed) was subtracted from the measured spectra.}
    \label{fig:background}
\end{figure}

\begin{figure}
    \centering
    \includegraphics[width=0.8\textwidth]{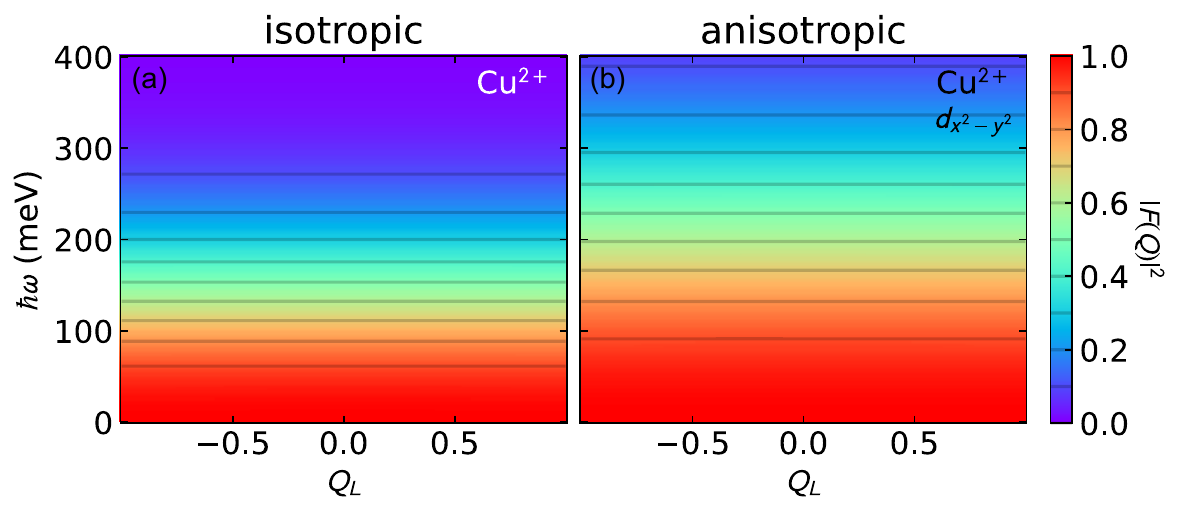}
    \caption{Results are shown assuming (a) isotropic Cu$^{2+}$ ions or (b) $d_{x^2-y^2}$ orbital order. There is a substantial difference between the two, 
    especially at high energies. The $d_{x^2-y^2}$ form factor was divided by $\rm Sr_{2.5} Ca_{11.5} Cu_{24} O_{41}$ scattering data in this study to obtain $S(q,\omega)$.}
    \label{fig:formfactor}
\end{figure}

\begin{figure}
    \centering
    \includegraphics[width=0.9\textwidth]{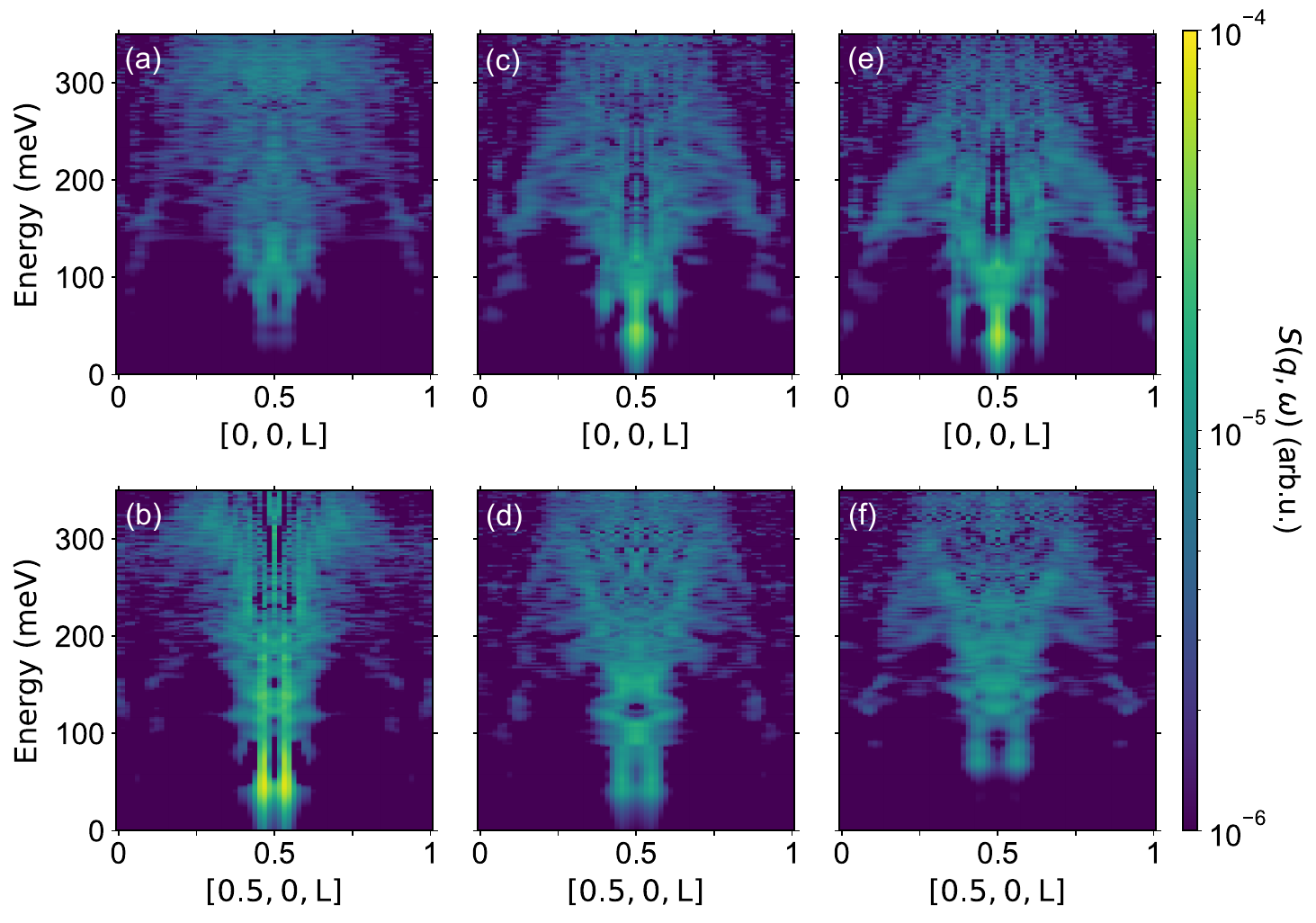}
    \caption{Dynamical spin structure factor for doped Hubbard model without extended interactions. The first column (panels a and b) shows $S({\bf q},\omega)$ results for $6.25\%$ doping along ${\bf q} = (q_\perp,0,L)$ with $q_\perp = 0$ and $\pi$, respectively. The second (c and d) and third (e and f) columns show the corresponding dynamical spin structure factors results for $9.375\%$ and $12.5\%$ doping, respectively. All results were obtained on $64\times 2$ site Hubbard ladders.}
    \label{fig:V0_Spectra}
\end{figure}

\clearpage
\section*{\bf Supplementary Note 1: Additional DMRG calculations}
To quantify the effect of $V$ on ladder doping, we estimate the effective bulk hole concentration using two different methods. 
Our first approach fits the density oscillations in the bulk region \cite{Dolfi_2015, PhysRevB.65.165122} using
\begin{equation}
     \langle n_j\rangle  =  A \cos \left( 2 \pi N_h j/L_\mathrm{bulk} + \phi_1 \right) + n_0
\end{equation}
where $\langle n_j \rangle$ is the total electron occupation on rung $j$, and the sum is restricted to a subset of sites where bulk-like density modulations are observed. ($L_\mathrm{bulk}$ is the number of rungs included in this region.) Our second approach estimates the doping by simply counting the average excess hole occupation $\delta = 1 - \frac{1}{2L_\mathrm{bulk}}\sum_j  \langle n_j \rangle $ in the same regions. The estimated effective doping levels obtained with these two methods are reported in Supplementary Figure ~\ref{fig:Doping_Estimates}. Both approaches produce the same qualitative picture, with slight differences in the quantitative values of effective doping. 

\begin{figure}[b]
    \centering
    \includegraphics[width=1\textwidth]{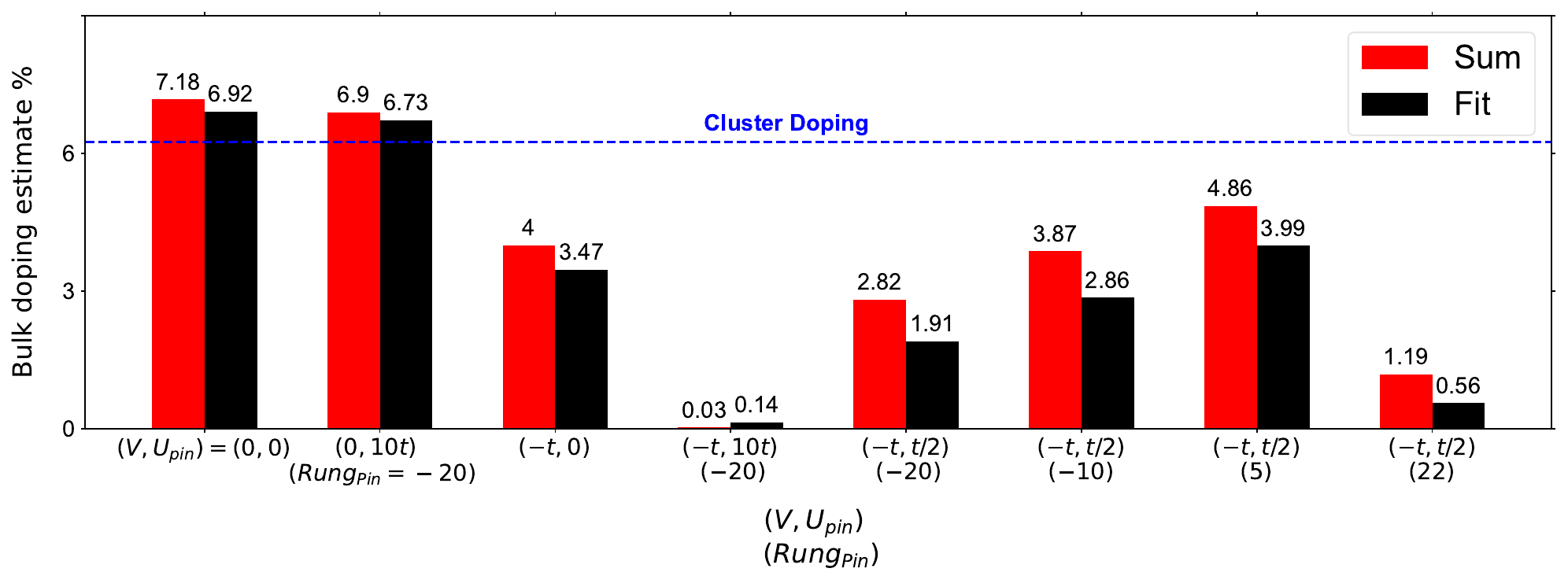}
    \caption{Bulk doping estimates for the various cluster calculations. The $Rung_{Pin}$ is the location of the pinning potential $U_{pin}$ when included. The bulk doping decreases when $V$ is present.}
    \label{fig:Doping_Estimates}
\end{figure}

For $V=0$, a strong potential $U_{pin}=10t$ localizes two holes at the impurity, leaving the hole density on the remainder of the ladder (including the edges) nearly unchanged. In contrast, introducing $V=-t$ reduces the bulk doping by almost half and collects the holes near any pinning center in the system. For a clean system, the open boundary conditions on the cluster behave as pinning centers. If we introduce additional pinning sites, they will also collect holes similar to the open edges of the cluster. Note that this occurs for strong and weak pinning potential placed anywhere in the system, as shown in Fig. 4 of the main text. 

\begin{figure}[t]
    \centering
    \includegraphics[width=\textwidth]{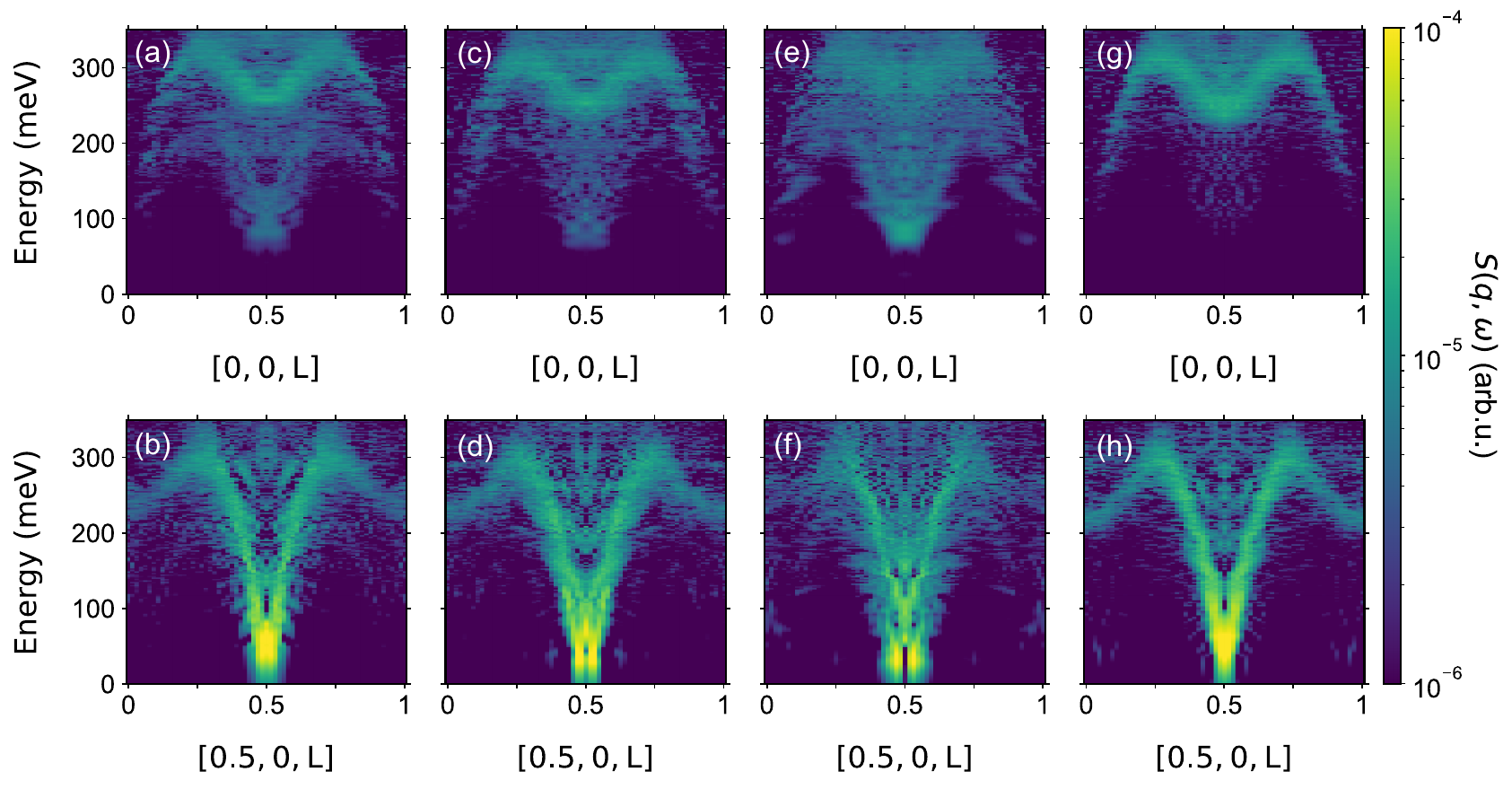}
    \caption{Dynamical spin structure factor for doped extended Hubbard model with a weak potential placed on different rungs. 
    The top row shows results for $q_\perp = 0$ with a weak potential $U_\mathrm{pin} = t/2$ placed on rungs (a) $-20$, (c) $-10$, (e) $5$, and (g) $22$ of the cluster. The bottom row (panels b, d, f, and h) shows the corresponding spectra for $q_\perp = \pi$. The main text also reports the case with pinning on rung $-20$.}
    \label{fig:SpectrawithPinning}
\end{figure}

\begin{figure*}[b]
	\centering\includegraphics[width=0.85\textwidth]{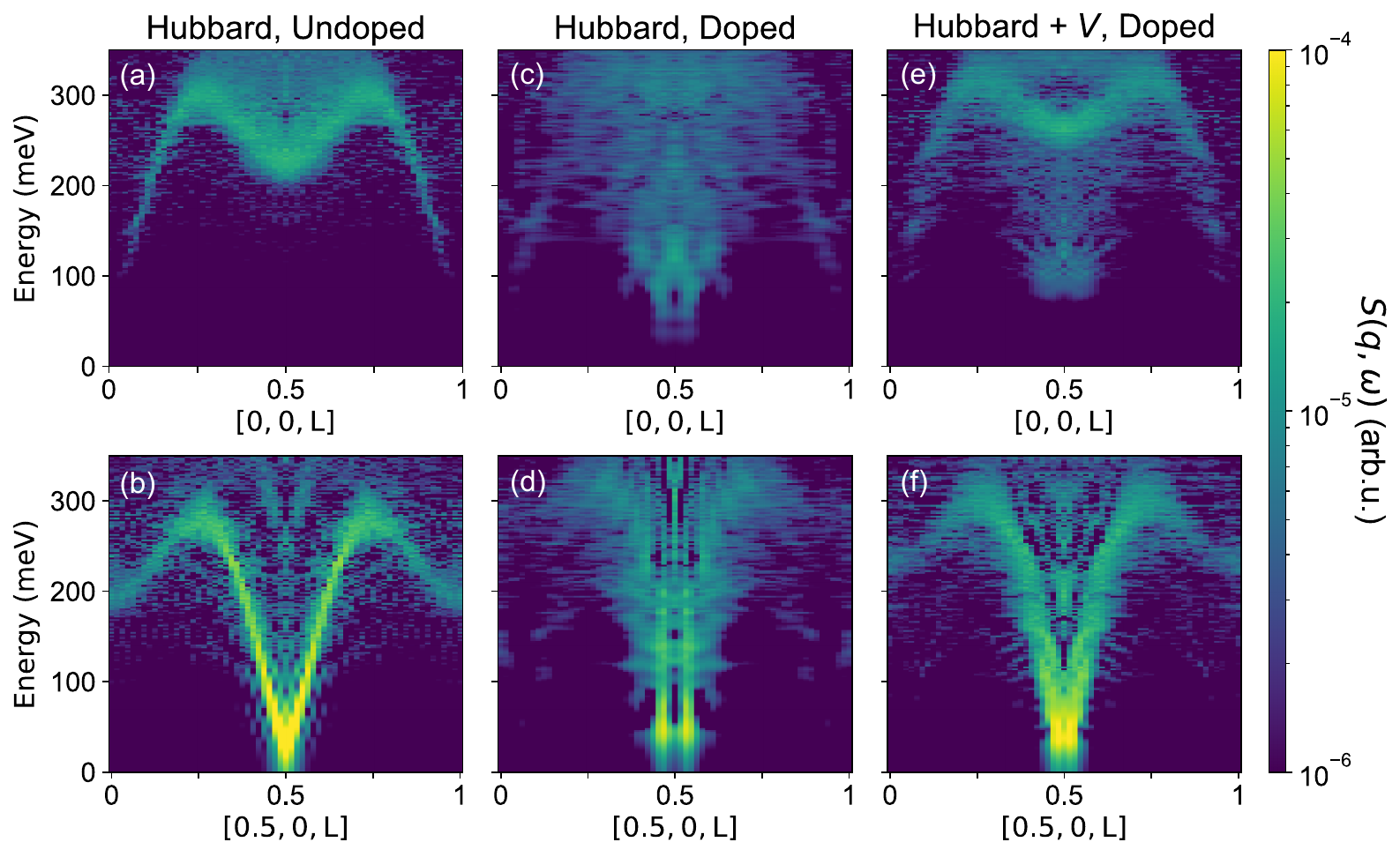}
	\caption{The calculated dynamical spin structure factor without instrumental broadening. (a), (b) $S({\bf q},\omega)$ for an undoped Hubbard Hamiltonian without any nearest-neighbor interactions ($V = 0$) along ${\bf q} = (q_\perp,0,L)$ with $q_\perp = 0$ and $\pi$, respectively. 
    (c), (d) $S({\bf q},\omega)$ for a $\rho = 6.25\%$ hole-doped Hubbard Hamiltonian, again with $V = 0$. 
    (e), (f) $S({\bf q},\omega)$ for a $\rho = 6.25\%$ hole-doped Hubbard Hamiltonian with an attractive nearest-neighbor interaction $V = -t$. All results were obtained with DMRG on a $64 \times 2$ ladder and have not been broadened with the instrumental resolution. }
	\label{flo:DMRG_SQW_compare}
\end{figure*}

Supplementary Figure \ref{fig:SpectrawithPinning} shows the $S({\bf q},\omega)$ spectra for the four weak impurity configurations used in Fig. 4 of the main text. The trend in doping estimates agrees with the locations of the incommensurate structure in the dynamical spin structure factor measured along $q_\perp = \pi$; larger doping corresponds to a larger separation in the incommensurate peaks.

For reference, Supplementary Figure \ref{flo:DMRG_SQW_compare} shows the same spectra as in Fig. 2 of the main text but without the convolution using the estimated momentum resolution function.  

\section{Binding Energy Calculations}
We define the binding energy as $\Delta_{B}(\delta)=E_2(\delta) +E_0(\delta)-2E_1(\delta)$, where $E_n(\delta)$ denotes the ground state energy for a system with $n$ holes introduced on top of a background with hole density $\delta$. We also define the spin gap $\Delta_{s}(\delta)=E(\delta, S^z_\mathrm{tot}=1)-E(\delta, S^z_\mathrm{tot}=0)$, where $E(\delta, S^z_\mathrm{tot})$ denotes the lowest energy for a system with hole density $\delta$ in magnetization sector $S^z_\mathrm{tot}$. These quantities are plotted as a function of system size in Supplementary Figure~\ref{flo:DMRGgaps} for the single-band extended Hubbard Hamiltonian, Eq. (1) in the main text, with $V=0$ and other parameters as given in the main text. Up to $m=5000$ DMRG states were kept in the calculations, yielding truncation errors below $10^{-6}$.

\begin{figure}[h]
\centering\includegraphics[width=0.6\columnwidth]{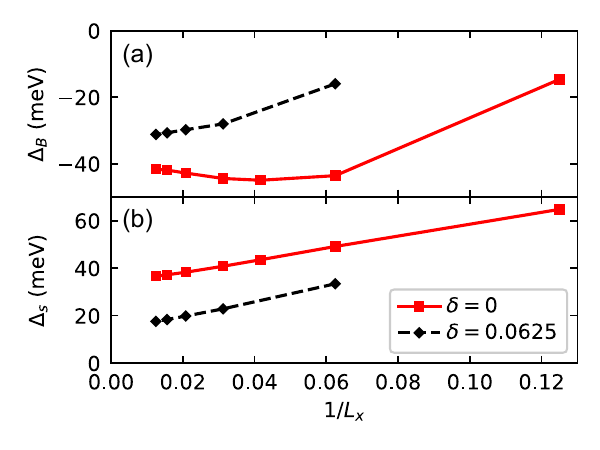}
 	\caption{Finite-size scaled energy gaps in the $V=0$ single-band extended Hubbard Hamiltonian calculated with DMRG. (a) the binding energy $\Delta_{B}$ as a function of $1/L_x$ for a $L_x \times 2$ ladder at half-filling ($\delta=0$) and finite doping ($\delta=0.0625$). The binding energy is negative, indicating the presence of two-hole bound states. (b) the spin gap $\Delta_{s}$ at half-filling and at finite doping.}
\label{flo:DMRGgaps}
\end{figure}

\bibliography{references}